\documentstyle[12pt,eepic,epic]{article}
\begin{document}
\begin{center}
\hfill HUPD-9716\\
\hfill Sept 13, 1997\\
\end{center}
\vspace{1mm}
\begin{center}
{\Large \bf Finite Grand Unified Theories and Inflation}
\end{center}
\vspace{2mm}
\begin{center}
\renewcommand{\thefootnote}{\fnsymbol{footnote}}
{\large S. Mukaigawa, T. Muta and S. D. Odintsov\footnote[2]{On leave of
absence from Department of Physics, Tomsk Pedagogical Institute, 634041
Tomsk, Russia}} \\
\vspace{5mm}
{\large Department of Physics, Hiroshima University \\
Higashi-Hiroshima, Hiroshima 739}
\end{center}
\begin{abstract}
A class of finite GUTs in curved spacetime is considered in connection
with the cosmological inflation scenario. It is confirmed that the use of 
the running scalar-gravitational coupling constant in these models helps
realizing a successful chaotic inflation. The analyses are made for some 
different sets of the models.
\end{abstract}

\section{Introduction}

It is now a common understanding to assume the presence of the inflationary 
stage in the early universe (see Ref. 1 for a review).
Among various models in the inflationary scenario the chaotic inflation
model seems to be the most successful and promising.$^{1}$
In the chaotic inflation model, however, we need the fine-tuning of some
 coupling constants such as scalar-gravitational coupling constant$^{2}$
$\xi$.
 The scalar-gravitational term associated with this coupling constant is
required in any quantum field theory in curved space-time in order to
 guarantee the multiplicative renormalizability of the theory$^{3}$
 (see Ref. 4 for a general review).
Applying the renormalization group argument we find that the coupling
 constant $\xi$ starts running.$^{3,4,5}$ The behavior of the running
coupling constant at strong gravitational field has been investigated for
 various models in Refs. 3 and 5 (see Ref. 4 for a review).

In a recent paper$^{6}$ an interesting observation has been made on an
implication of the running coupling constant in realizing the inflation
 scenario. The authors of Ref. 6 report that the use of the running
scalar-gravitational coupling constant in a specific field theory$^{7}$
 helps constructing a successful model of the chaotic inflation.
The behaviour of the running $\xi$ in their model is typical of
 $\lambda\varphi^{4}$-theory $^{3}$ so that
\begin{eqnarray}
\xi(t) &=& \frac{1}{6} + (\xi-\frac{1}{6})(1-a^{2}\lambda t)^{\alpha},
\nonumber \\
\lambda(t) &=& \frac{\lambda}{1-a^2 \lambda t},
\label{eq1}
\end{eqnarray}
where $a^2$ and $\alpha$ are suitable constants, and $t$ is the RG
parameter,
$ t=\frac{1}{2} \ln \frac{\varphi^2}{\mu^2}$ with $\mu$ the
renormalization scale. Depending on the sign of the exponent $\alpha$
 it is possible to have an asymptotic conformal invariance or 
not.$^{3,4,5}$
 In fact $\xi(t)\rightarrow 1/6$ (for $\alpha<0$) as $t \rightarrow 
\infty$
 (the infrared limit ) which is the case of Ref. 6.
 For $\alpha > 0, |\xi(t)| \rightarrow \infty$ as $t \rightarrow
\infty$.$^{3,4,5}$ In Ref. 6 one of the simplest supersymmetric model,
 i.e. the Wess-Zumino model, was taken into account and the effective
 potential in the flat direction of the model was considered in order to
 restrict oneself to the effect of quadratic terms of the model.
Thus the model exhibits the typical behavior as mentioned above.

There is an another type of supersymmetric models which are called finite
GUTs in which the behaviour of $\xi(t)$ is qualitatively different from the 
one which we have seen in Eq. (\ref{eq1}).
The purpose of this note is to examine a possibility
of the chaotic inflation in the finite GUTs (including some possible
finite non-supersymmetric theories).

\section{Finite GUTs}

Let us consider the typical (supersymmetric or non-supersymmetric) finite
GUTs in curved spacetime.
In curved space-time the theory is not completely finite$^{8}$
 according to an appearence of the divergence in the vacuum energy
(in the external gravitational field sector).
Nevertheless, the matter sector remains unaffected.$^{8}$
 The behavior of coupling constants in the matter sector is given
 as follows,$^{8}$
\begin{eqnarray}
&&g^{2}(t)=g^2, \;\;\;  h(t)^{2}=k_{1}g^{2}, \;\;\;  f(t)=k_{2}g^2, 
\nonumber  \\
&& \xi(t) = \frac{1}{6}+ ( \xi - \frac{1}{6})\exp \left[ c g^{2} t \right],
\label{eq2}
\end{eqnarray}
where $g(t)$, $h(t)$ and $f(t)$ are running coupling constants respectively,
$g_{2} \ll 1$, and $k_{1}, k_{2}$ and $c$ are certain numerical constants
 determined by the group structure of the theory.
Note that $c$ may be positive,
negative or zero depending on the nature of the theory.
Let us consider a concrete example.
The $SU(2)$ gauge theory with $SU(N)$ global
 invariance (the Lagrangian is written in the book previously
mentioned;$^{4}$ see Eq. (3.130) therein.) includes gauge fields,
Weyl spinors and scalars in the adjoint representation of $SU(2)$.
With respect to the global group $SU(N)$ the spinors and scalars belong to 
the fundamental representation and six-dimensional antisymmetric adjoint
representation, respectively.
In the case of flat space-time the theory has been introduced in Ref. 9.

The theory has two regimes.
In the first regime it is $N=4$ extended supersymmetric
 gauge theory which is finite to all orders of perturbation theory in flat 
spacetime.The direct calculation yields$^{8}$ $c={3}/{(2\pi^2)}$.
In the second regime the theory corresponds to the one-loop finite
non-supersymmetric theory in flat spacetime and $c \approx
{27}/{(4\pi)^2}$.
There are some other finite theories where $c$ could be negative or
zero.$^{8}$

Consider the scalar sector of the finite theory taken in the flat
direction of the effective potential where interaction terms do not
contribute. The renormalization-group-improved effective action in curved
 spacetime $^{4,10}$ coupled with the classical Einstein gravity is given 
by
\begin{equation}
S= - \frac{1}{2} \int d^{4}x \sqrt{-g} \left[ \frac{M_{pl}}{8\pi} R
 +Z(t) \partial_{\mu}\varphi(t) \partial^{\mu}\varphi(t)
 +m^{2}(t)\varphi^{2}(t) -\xi(t) R \varphi^{2}(t) \right].
\end{equation}
where $t=\frac{1}{2} \ln{ \frac{\varphi^{2}}{\mu^{2}} }$.
 The variation of the running mass as a function of $t$ is considered to 
be small, i.e. $m^{2}(t) \simeq m^{2}$.
We choose the gauge in which the anomalous dimension for the scalar field
vanishes (i.e. the renormalization for the scalar field is finite), then
\begin{equation}
Z(t) = 1,  \;\;\;  \varphi(t) = \varphi.
\label{eq4}
\end{equation}
Under this circumstance the effective action reads
\begin{equation}
S=-\frac{1}{2} \int d^{4}x \sqrt{-g}
\left[ \frac{M_{pl}}{8\pi} R
 + \partial_{\mu}\varphi \partial^{\mu}\varphi
+ m^{2} \varphi^{2} -\xi(t) R \varphi^{2}
\right].
\label{eq5}
\end{equation}
It should be noted here that in our present model it is not necessary to
incorporate the effect of the running $\varphi$ due to the property (4).
In the model employed in Ref. 6 Eq. (\ref{eq4}) does not hold so that
the effect of the running $\varphi$ plays an important role in deriving
the cosmological predictions while in Ref. 6 this effect is not fully
taken into account.

Field equations for the theory characterized by Eq. (\ref{eq5}) are given
by
\begin{eqnarray}
&& \left( \Box - m^{2} + \xi{\varphi}R +
  \frac{1}{2}\frac{d\xi}{d\varphi}(\varphi) R \right) \varphi
\;\; = \; 0, \nonumber \\
&& R_{\mu\nu} -\frac{1}{2}R g_{\mu\nu} \;\; = \;\;
\frac{8\pi}{M_{pl}} T_{\mu\nu}.
\end{eqnarray}
Rewriting these equations in the Friedmann-Robertson-Walker Universe
 with scale factor $a$  we obtain$^{11}$
\begin{eqnarray}
&& \ddot{\varphi}+3H\dot{\varphi}+m^{2}\varphi
+ \left[
\xi (\varphi) \varphi + \frac{1}{2} \frac{ d \xi }{ d \varphi } (\varphi)
\right]
\times \nonumber \\
&& \left[
\{
6 \xi (\varphi) -1 + 12 \frac{ d \xi }{ d \varphi } (\varphi) \varphi
+ 6 \frac{ d^{2} \xi }{ d \varphi^{2} } (\varphi) \varphi^{2}
\}
\right.
\dot{ \varphi }^{2} + 2m_{2}\varphi^{2} \\
&& \left.
+ \{ 6 \xi (\varphi) \varphi + 3 \frac{d\xi}{d\varphi}(\varphi)\varphi^{2} 
\}
\{ \ddot{\varphi} + 3H\dot{\varphi} \}
\right]
\left[
\frac{M_{pl}}{8\pi} - \xi(\varphi)\varphi^{2}
\right]^{-1} = 0, \nonumber
\end{eqnarray}
where $H={\dot{a}}/{a}$.

\section{Chaotic inflation}

As it has been established in Refs. 11 there are two saddle points
 of Eq. (7) for negative $\xi$. They are given by
\begin{equation}
\varphi_{\mbox{cr}} = \pm \frac{ M_{pl} }{ \sqrt{-8\pi\xi} }, \;\;\;
\dot{\varphi}=0.
\label{eq8}
\end{equation}

It is discussed in Ref. 6 (and in the preceding works cited there) that
the initial conditions for inflaton $\varphi$ in order to have a 
successful
 chaotic inflation as well as a sufficient period of the inflation are

\begin{equation}
-\frac{M_{pl}}{\sqrt{8\pi |\xi|}} < \varphi < \frac{M_{pl}}{\sqrt{8\pi
|\xi|}},
\label{eq9}
\end{equation}
and
\begin{equation}
|\varphi| \geq 5 M_{pl}.
\label{eq10}
\end{equation}
Actually for negative as well as positive $\xi$ we have two qualitatively
 different situations summarrized by the condition (\ref{eq9}) and
(\ref{eq10})

Our purpose now is to examine initial conditions (\ref{eq9}) and
(\ref{eq10})
for finite GUTs with running $\xi(t)$ given by Eq. (\ref{eq2}).
It should be noted that in Eq. (\ref{eq2}) $\xi = \xi(\mu)$ and $g^2 =
g^2(\mu)$
 are initial values of the running coupling constants $\xi(t)$ and
$g^2(t)$
 respectively at a certain RG scale $\mu$ .
We start with a very small initial value of $\xi(t)$ and so we may
practically
set $\xi(\mu)=0$. Then the total variation of $\xi(t)$ comes purely from
the running effect. We start with the minimal theory at scale $\mu$.
We wish to plot $\varphi_{\mbox{cr}}-\varphi$ as a function of $\varphi$.
The relation
between $\varphi_{\mbox{cr}}-\varphi$ and $\varphi$ is easily obtained by
using
Eq. (\ref{eq2}) and Eq. (\ref{eq8}) with $ t=\frac{1}{2} \ln
\frac{\varphi^2}{\mu^2}$.
In Figs.1, 2, 3 and 4 the behavior of $\varphi_{\mbox{cr}}-\varphi$ is shown 
in four typical cases with ($\mu=50M_{pl}, c<0$), ($\mu=50M_{pl}, c>0$),
($\mu=2M_{pl}, c<0$)
and ($\mu=2M_{pl}, c>0$) respectively. (Note here that the behavior of
$\varphi_{\mbox{cr}}-\varphi$ as a function of $\varphi$ is symmetric
around
$\varphi=0$ and so we need not to consider the case $\varphi<0$).\\

\begin{center}
Figs. 1, 2, 3 and 4 \\
\end{center}

Let us examine whether we can have any region where
$\varphi_{\mbox{cr}}-\varphi$ is
 positive (i. e. $\varphi_{\mbox{cr}}>\varphi$) for small $\varphi$ starting 
with $\varphi\ge 5M_{pl}$. We clearly see that in all four cases
$\varphi_{\mbox{cr}}-\varphi$
 becomes positive if $|c|g^2\sim 10^{-3}$. It is important to note that the 
chaotic inflation is realized independent of the sign of $c$. Thus we
 conclude that for a wide class of the finite GUTs we have successful
 chaotic inflations.

\section{Conclusions}

Working within the framework of the finite GUTs we examined the
mechanism$^{6}$ that may lead to the successful chaotic inflation.
We find that in a wide class of the
finite GUTs the chaotic inflation is realized as far as the gauge coupling 
constant is kept sufficiently small.
In this sense the finite GUTs are worth for further
 investigations in connection with the early universe scenario.

In this regards it is very interesting to note that the finite GUTs in 
curved spacetime are one of the possible canditates to give a solution 
to the cosmological constant problem
 (see the second reference in Ref. 10) due to the exponential running of the 
effective cosmological constant.
These faborable properties of the theories under discussion indicate that
the cosmological applications of the finite GUTs (in particular the N=4 
super Yang-Mills theory) should be considered more seriously.
\pagebreak

\begin{flushleft}
{\Large \bf References}
\end{flushleft}
\begin{enumerate}
\item
A. Linde,
{\it Particle Physics and Inflationary Cosmology},
Harwood Academic Publishers,1990.

\item
T. Futamase, T. Rothman and R. Matzner,
Phys. Rev. \underline{D39} (1989) 405;
T. Futamase and K. Maeda, Phys.Rev. \underline{D39} (1989) 405.

\item
I. L. Buchbinder and S. D. Odintsov,
Izw. Vuzov. Fizika (Sov. Phys. J.) N12 (1983) 108;
Yad. Fiz. (Sov. J. Nucl. Phys.) \underline{40} (1984) 1338;
Lett. Nuovo Cim. \underline{42} (1985) 379.

\item
I. L. Buchbinder, S. D. Odintsov and I. L. Shapiro,
{\it Effective Action in Quantum Gravity},
IOP Publishing, Bristol and Philadelphia, 1992.

\item
T. Muta and S. D. Odintsov, Mod. Phys. Lett. A6 (1991) 3641.

\item
T. Futamase and M. Tanaka,
preprint OCHA-PP-95, 1997;
hep-ph/9704303.

\item
M. Tanaka,
hep-th/9701063.

\item
I. L. Buchbinder, S. D. Odintsov and I. M. Lichtzier,
Class. Quant. Grav. \underline{6} (1989) 605.

\item
M.B\"ohm and A. Denner,
Nucl. Phys. \underline{B282} (1987) 206.

\item
I. L. Buchbinder, S. D. Odintsov,
Class. Quant. Grav. \underline{2} (1985) 721;
E. Elizalde and S. D. Odintsov, Phys. Lett. B333 (1994) 331.

\item
L. Amendola, M. Litterio and F. Occhionero,
Int. J. Mod. Phys. \underline{A5} (1990) 3861;
A. Barroso, J. Casasayas, P. Crawford, P. Moniz and A. Nunes,
Phys.Lett. \underline{B275}(1992)264.

\end{enumerate}
\pagebreak
\begin{figure}
\setlength{\unitlength}{0.240900pt}
\begin{picture}(1500,900)(0,0)
\tenrm
\thinlines \drawline[-50](220,562)(1436,562)
\thicklines \path(220,113)(240,113)
\thicklines \path(1436,113)(1416,113)
\put(198,113){\makebox(0,0)[r]{-10}}
\thicklines \path(220,203)(240,203)
\thicklines \path(1436,203)(1416,203)
\put(198,203){\makebox(0,0)[r]{-8}}
\thicklines \path(220,293)(240,293)
\thicklines \path(1436,293)(1416,293)
\put(198,293){\makebox(0,0)[r]{-6}}
\thicklines \path(220,383)(240,383)
\thicklines \path(1436,383)(1416,383)
\put(198,383){\makebox(0,0)[r]{-4}}
\thicklines \path(220,473)(240,473)
\thicklines \path(1436,473)(1416,473)
\put(198,473){\makebox(0,0)[r]{-2}}
\thicklines \path(220,562)(240,562)
\thicklines \path(1436,562)(1416,562)
\put(198,562){\makebox(0,0)[r]{0}}
\thicklines \path(220,652)(240,652)
\thicklines \path(1436,652)(1416,652)
\put(198,652){\makebox(0,0)[r]{2}}
\thicklines \path(220,742)(240,742)
\thicklines \path(1436,742)(1416,742)
\put(198,742){\makebox(0,0)[r]{4}}
\thicklines \path(220,832)(240,832)
\thicklines \path(1436,832)(1416,832)
\put(198,832){\makebox(0,0)[r]{6}}
\thicklines \path(220,113)(220,133)
\thicklines \path(220,832)(220,812)
\put(220,68){\makebox(0,0){0}}
\thicklines \path(463,113)(463,133)
\thicklines \path(463,832)(463,812)
\put(463,68){\makebox(0,0){2}}
\thicklines \path(706,113)(706,133)
\thicklines \path(706,832)(706,812)
\put(706,68){\makebox(0,0){4}}
\thicklines \path(950,113)(950,133)
\thicklines \path(950,832)(950,812)
\put(950,68){\makebox(0,0){6}}
\thicklines \path(1193,113)(1193,133)
\thicklines \path(1193,832)(1193,812)
\put(1193,68){\makebox(0,0){8}}
\thicklines \path(1436,113)(1436,133)
\thicklines \path(1436,832)(1436,812)
\put(1436,68){\makebox(0,0){10}}
\thicklines \path(220,113)(1436,113)(1436,832)(220,832)(220,113)
\put(45,922){\makebox(0,0)[l]{\shortstack{$(\varphi_{cr}-\varphi)/M_{pl}$}}}
\put(828,-22){\makebox(0,0){$\varphi/M_{pl}$}}
\put(828,877){\makebox(0,0){$\mu/M_{pl} = 50$ , $c < 0$}}
\put(950,652){\makebox(0,0)[l]{$-cg^2 = 10^{-3}$}}
\put(950,428){\makebox(0,0)[l]{$-cg^2 = 10^{-2}$}}
\put(706,293){\makebox(0,0)[l]{$-cg^2 = 10^{-1}$}}
\thinlines \path(220,570)(220,570)(221,572)(222,573)(222,573)(223,573)
(224,573)(225,573)(226,573)(226,573)(228,573)(229,572)(233,572)(245,569)
(270,561)(295,553)(319,545)(344,537)(369,528)(394,520)(419,511)(443,502)
(468,494)(493,485)(518,476)(543,468)(568,459)(592,450)(617,441)(642,433)
(667,424)(692,415)(716,406)(741,397)(766,388)(791,380)(816,371)(840,362)
(865,353)(890,344)(915,335)(940,326)(965,318)(989,309)(1014,300)(1039,291)
(1064,282)(1089,273)(1113,264)(1138,255)(1163,247)
\thinlines \path(1163,247)(1188,238)(1213,229)(1237,220)(1262,211)
(1287,202)(1312,193)(1337,184)(1362,175)(1386,166)(1411,158)(1436,149)
\thinlines \path(220,607)(220,607)(221,612)(222,614)(222,615)(223,616)
(224,616)(225,617)(226,617)(226,617)(227,617)(228,618)(229,618)(229,618)
(230,618)(231,618)(232,618)(233,618)(233,618)(234,618)(235,618)(236,618)
(236,618)(237,618)(239,618)(240,618)(242,618)(245,618)(251,617)(257,616)
(270,613)(295,607)(319,601)(344,594)(369,586)(394,579)(419,571)(443,564)
(468,556)(493,548)(518,540)(543,532)(568,524)(592,516)(617,508)(642,500)
(667,492)(692,484)(716,476)(741,468)(766,459)
\thinlines \path(766,459)(791,451)(816,443)(840,435)(865,426)(890,418)
(915,410)(940,402)(965,393)(989,385)(1014,377)(1039,368)(1064,360)
(1089,352)(1113,343)(1138,335)(1163,327)(1188,318)(1213,310)(1237,301)
(1262,293)(1287,285)(1312,276)(1337,268)(1362,260)(1386,251)(1411,243)
(1436,234)
\thinlines \path(220,711)(220,711)(221,727)(222,733)(222,736)(223,739)
(225,744)(226,747)(229,751)(233,755)(236,757)(239,759)(242,761)(245,762)
(251,764)(257,766)(260,766)(264,767)(267,767)(270,767)(273,768)(276,768)
(279,768)(281,768)(282,768)(284,768)(285,768)(287,768)(288,768)(288,768)
(289,768)(290,768)(291,768)(291,768)(292,768)(293,768)(294,768)(295,768)
(296,768)(298,768)(301,768)(304,768)(307,768)(319,767)(332,766)(344,765)
(369,762)(394,758)(419,754)(443,749)(468,745)
\thinlines \path(468,745)(493,740)(518,735)(543,729)(568,724)(592,718)
(617,712)(642,707)(667,701)(692,695)(716,689)(741,683)(766,676)(791,670)
(816,664)(840,658)(865,651)(890,645)(915,639)(940,632)(965,626)(989,619)
(1014,613)(1039,606)(1064,600)(1089,593)(1113,586)(1138,580)(1163,573)
(1188,567)(1213,560)(1237,553)(1262,547)(1287,540)(1312,533)(1337,527)
(1362,520)(1386,513)(1411,506)(1436,500)
\end{picture}
\caption{Behavior of $\varphi_{\mbox{cr}}-\varphi$ as a function of 
$\varphi$ for
$\mu=50M_{pl}$ and $c<0$. The three typical cases of the gauge coupling 
strength are
shown: $-cg^2=10^{-3},10^{-2},10^{-1}$.}
\end{figure}
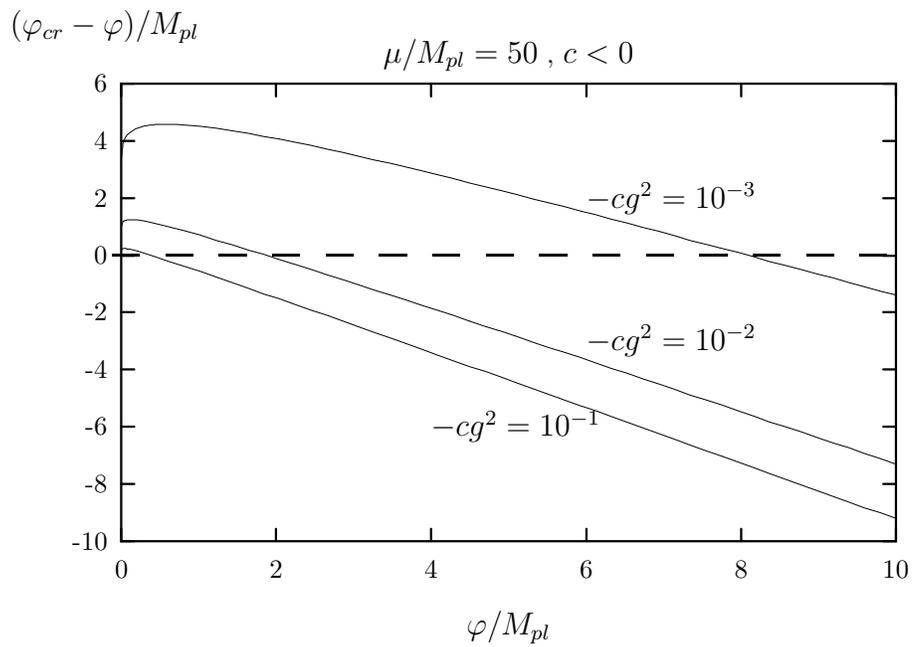
\pagebreak
\begin{figure}
\setlength{\unitlength}{0.240900pt}
\begin{picture}(1500,900)(0,0)
\tenrm
\thinlines \drawline[-50](220,562)(1436,562)
\thicklines \path(220,113)(240,113)
\thicklines \path(1436,113)(1416,113)
\put(198,113){\makebox(0,0)[r]{-10}}
\thicklines \path(220,203)(240,203)
\thicklines \path(1436,203)(1416,203)
\put(198,203){\makebox(0,0)[r]{-8}}
\thicklines \path(220,293)(240,293)
\thicklines \path(1436,293)(1416,293)
\put(198,293){\makebox(0,0)[r]{-6}}
\thicklines \path(220,383)(240,383)
\thicklines \path(1436,383)(1416,383)
\put(198,383){\makebox(0,0)[r]{-4}}
\thicklines \path(220,473)(240,473)
\thicklines \path(1436,473)(1416,473)
\put(198,473){\makebox(0,0)[r]{-2}}
\thicklines \path(220,562)(240,562)
\thicklines \path(1436,562)(1416,562)
\put(198,562){\makebox(0,0)[r]{0}}
\thicklines \path(220,652)(240,652)
\thicklines \path(1436,652)(1416,652)
\put(198,652){\makebox(0,0)[r]{2}}
\thicklines \path(220,742)(240,742)
\thicklines \path(1436,742)(1416,742)
\put(198,742){\makebox(0,0)[r]{4}}
\thicklines \path(220,832)(240,832)
\thicklines \path(1436,832)(1416,832)
\put(198,832){\makebox(0,0)[r]{6}}
\thicklines \path(220,113)(220,133)
\thicklines \path(220,832)(220,812)
\put(220,68){\makebox(0,0){0}}
\thicklines \path(463,113)(463,133)
\thicklines \path(463,832)(463,812)
\put(463,68){\makebox(0,0){2}}
\thicklines \path(706,113)(706,133)
\thicklines \path(706,832)(706,812)
\put(706,68){\makebox(0,0){4}}
\thicklines \path(950,113)(950,133)
\thicklines \path(950,832)(950,812)
\put(950,68){\makebox(0,0){6}}
\thicklines \path(1193,113)(1193,133)
\thicklines \path(1193,832)(1193,812)
\put(1193,68){\makebox(0,0){8}}
\thicklines \path(1436,113)(1436,133)
\thicklines \path(1436,832)(1436,812)
\put(1436,68){\makebox(0,0){10}}
\thicklines \path(220,113)(1436,113)(1436,832)(220,832)(220,113)
\put(45,922){\makebox(0,0)[l]{\shortstack{$(\varphi_{cr}-\varphi)/M_{pl}$}}}
\put(828,-22){\makebox(0,0){$\varphi/M_{pl}$}}
\put(828,877){\makebox(0,0){$\mu/M_{pl} = 50$ , $c > 0$}}
\put(950,652){\makebox(0,0)[l]{$cg^2 = 10^{-3}$}}
\put(950,428){\makebox(0,0)[l]{$cg^2 = 10^{-2}$}}
\put(706,293){\makebox(0,0)[l]{$cg^2 = 10^{-1}$}}
\thinlines \path(220,586)(220,586)(221,586)(222,586)(222,586)
(223,586)(225,586)(226,585)(233,584)(245,580)(270,572)(295,564)
(319,555)(344,546)(369,538)(394,529)(419,520)(443,511)(468,503)
(493,494)(518,485)(543,476)(568,467)(592,458)(617,449)(642,441)
(667,432)(692,423)(716,414)(741,405)(766,396)(791,387)(816,378)
(840,369)(865,360)(890,352)(915,343)(940,334)(965,325)(989,316)
(1014,307)(1039,298)(1064,289)(1089,280)(1113,271)(1138,262)(1163,253)
(1188,244)(1213,235)(1237,226)(1262,217)
\thinlines \path(1262,217)(1287,209)(1312,200)(1337,191)(1362,182)
(1386,173)(1411,164)(1436,155)
\thinlines \path(220,612)(220,612)(221,617)(222,618)(222,619)
(223,620)(224,620)(225,621)(226,621)(226,621)(227,622)(228,622)
(229,622)(229,622)(230,622)(231,622)(232,622)(233,622)(233,622)
(234,622)(235,622)(236,622)(237,622)(239,622)(242,622)(245,621)
(257,619)(270,617)(295,610)(319,604)(344,597)(369,589)(394,582)
(419,574)(443,567)(468,559)(493,551)(518,543)(543,535)(568,527)
(592,519)(617,511)(642,503)(667,495)(692,486)(716,478)(741,470)
(766,462)(791,454)(816,445)(840,437)
\thinlines \path(840,437)(865,429)(890,420)(915,412)(940,404)
(965,395)(989,387)(1014,379)(1039,370)(1064,362)(1089,354)(1113,345)
(1138,337)(1163,329)(1188,320)(1213,312)(1237,304)(1262,295)(1287,287)
(1312,278)(1337,270)(1362,262)(1386,253)(1411,245)(1436,236)
\thinlines \path(220,712)(220,712)(221,728)(222,734)(222,738)
(223,741)(225,745)(226,748)(229,753)(233,756)(236,758)(239,760)
(242,762)(245,763)(251,765)(257,767)(260,767)(264,768)(267,768)
(270,768)(273,769)(276,769)(279,769)(281,769)(282,769)(284,769)
(285,769)(287,769)(288,769)(288,769)(289,769)(290,769)(291,769)
(291,769)(292,769)(293,769)(294,769)(295,769)(296,769)(298,769)
(301,769)(304,769)(307,769)(319,768)(332,767)(344,766)(369,763)
(394,759)(419,755)(443,750)(468,746)
\thinlines \path(468,746)(493,741)(518,735)(543,730)(568,725)
(592,719)(617,713)(642,707)(667,702)(692,696)(716,689)(741,683)
(766,677)(791,671)(816,665)(840,658)(865,652)(890,646)(915,639)
(940,633)(965,626)(989,620)(1014,613)(1039,607)(1064,600)(1089,594)
(1113,587)(1138,580)(1163,574)(1188,567)(1213,561)(1237,554)(1262,547)
(1287,541)(1312,534)(1337,527)(1362,520)(1386,514)(1411,507)(1436,500)
\end{picture}
\caption{Behavior of $\varphi_{\mbox{cr}}-\varphi$ as a function of 
$\varphi$ for
$\mu=50M_{pl}$ and $c>0$. The three typical cases of the gauge coupling 
strength are
shown: $cg^2=10^{-3},10^{-2},10^{-1}$.}
\end{figure}
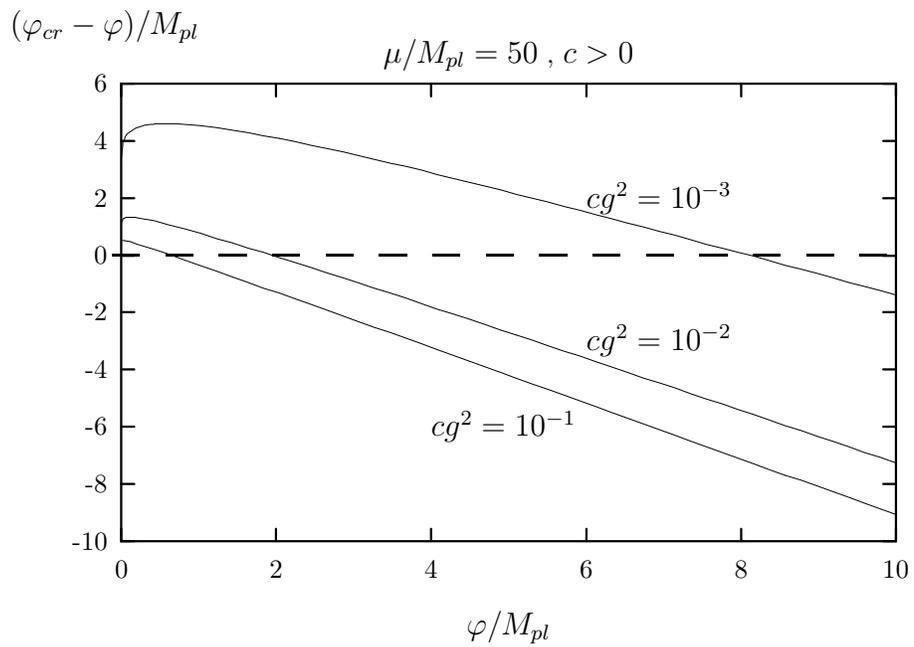
\pagebreak
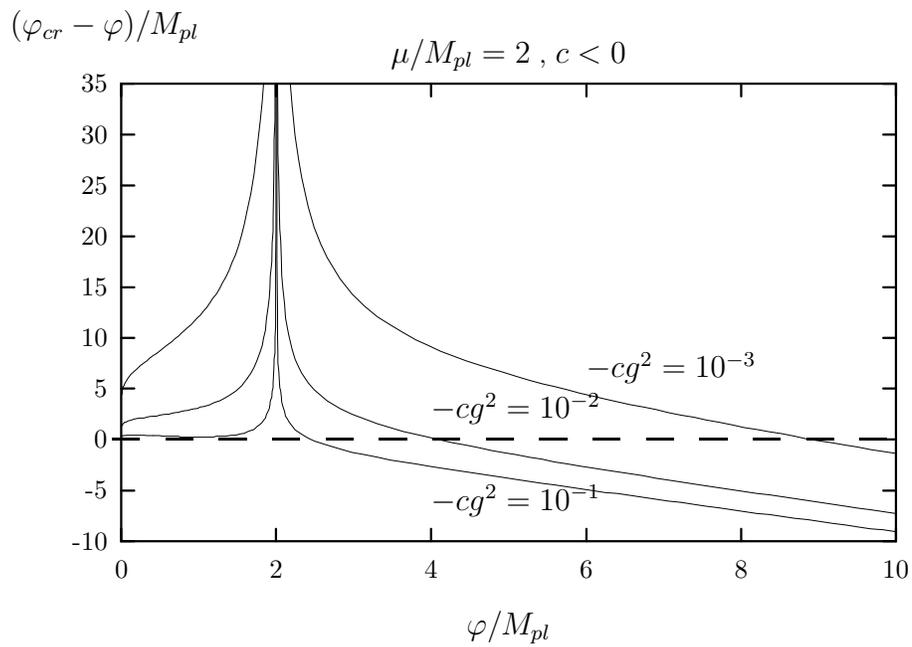
\begin{figure}
\setlength{\unitlength}{0.240900pt}
\begin{picture}(1500,900)(0,0)
\tenrm
\thinlines \drawline[-50](220,273)(1436,273)
\thicklines \path(220,113)(240,113)
\thicklines \path(1436,113)(1416,113)
\put(198,113){\makebox(0,0)[r]{-10}}
\thicklines \path(220,193)(240,193)
\thicklines \path(1436,193)(1416,193)
\put(198,193){\makebox(0,0)[r]{-5}}
\thicklines \path(220,273)(240,273)
\thicklines \path(1436,273)(1416,273)
\put(198,273){\makebox(0,0)[r]{0}}
\thicklines \path(220,353)(240,353)
\thicklines \path(1436,353)(1416,353)
\put(198,353){\makebox(0,0)[r]{5}}
\thicklines \path(220,433)(240,433)
\thicklines \path(1436,433)(1416,433)
\put(198,433){\makebox(0,0)[r]{10}}
\thicklines \path(220,512)(240,512)
\thicklines \path(1436,512)(1416,512)
\put(198,512){\makebox(0,0)[r]{15}}
\thicklines \path(220,592)(240,592)
\thicklines \path(1436,592)(1416,592)
\put(198,592){\makebox(0,0)[r]{20}}
\thicklines \path(220,672)(240,672)
\thicklines \path(1436,672)(1416,672)
\put(198,672){\makebox(0,0)[r]{25}}
\thicklines \path(220,752)(240,752)
\thicklines \path(1436,752)(1416,752)
\put(198,752){\makebox(0,0)[r]{30}}
\thicklines \path(220,832)(240,832)
\thicklines \path(1436,832)(1416,832)
\put(198,832){\makebox(0,0)[r]{35}}
\thicklines \path(220,113)(220,133)
\thicklines \path(220,832)(220,812)
\put(220,68){\makebox(0,0){0}}
\thicklines \path(463,113)(463,133)
\thicklines \path(463,832)(463,812)
\put(463,68){\makebox(0,0){2}}
\thicklines \path(706,113)(706,133)
\thicklines \path(706,832)(706,812)
\put(706,68){\makebox(0,0){4}}
\thicklines \path(950,113)(950,133)
\thicklines \path(950,832)(950,812)
\put(950,68){\makebox(0,0){6}}
\thicklines \path(1193,113)(1193,133)
\thicklines \path(1193,832)(1193,812)
\put(1193,68){\makebox(0,0){8}}
\thicklines \path(1436,113)(1436,133)
\thicklines \path(1436,832)(1436,812)
\put(1436,68){\makebox(0,0){10}}
\thicklines \path(220,113)(1436,113)(1436,832)(220,832)(220,113)
\put(45,922){\makebox(0,0)[l]{\shortstack{$(\varphi_{cr}-\varphi)/M_{pl}$}}}
\put(828,-22){\makebox(0,0){$\varphi/M_{pl}$}}
\put(828,877){\makebox(0,0){$\mu/M_{pl} = 2$ , $c < 0$}}
\put(950,385){\makebox(0,0)[l]{$-cg^2 = 10^{-3}$}}
\put(706,321){\makebox(0,0)[l]{$-cg^2 = 10^{-2}$}}
\put(706,177){\makebox(0,0)[l]{$-cg^2 = 10^{-1}$}}
\thinlines \path(464,832)(464,557)(465,474)(465,434)(466,410)
(466,393)(467,380)(468,362)(469,355)(470,349)(471,340)(472,332)
(473,326)(476,317)(478,310)(481,305)(483,301)(488,294)(493,288)
(503,281)(513,275)(523,271)(543,264)(563,258)(582,253)(602,249)
(622,245)(642,241)(662,238)(682,235)(702,231)(721,228)(741,225)
(761,222)(781,219)(801,216)(821,213)(840,210)(860,207)(880,204)
(900,202)(920,199)(940,196)(960,193)(979,190)(999,188)(1019,185
)(1039,182)(1059,179)(1079,176)(1099,174)
\thinlines \path(1099,174)(1118,171)(1138,168)(1158,166)(1178,163)
(1198,160)(1218,157)(1237,155)(1257,152)(1277,149)(1297,147)(1317,144)
(1337,141)(1357,139)(1376,136)(1396,133)(1416,131)(1436,128)
\thinlines \path(465,832)(466,776)(466,722)(467,681)(468,624)(469,603)
(470,585)(471,555)(472,532)(473,514)(476,486)(478,464)(481,448)
(483,435)(488,414)(493,399)(498,387)(503,377)(513,362)(523,351)
(543,334)(563,322)(582,313)(602,305)(622,298)(642,292)(662,286)
(682,281)(702,277)(721,272)(741,268)(761,264)(781,260)(801,256)
(821,252)(840,249)(860,245)(880,242)(900,238)(920,235)(940,231)
(960,228)(979,225)(999,222)(1019,219)(1039,216)(1059,212)(1079,209)
(1099,206)(1118,203)(1138,200)
\thinlines \path(1138,200)(1158,197)(1178,194)(1198,191)(1218,188)
(1237,186)(1257,183)(1277,180)(1297,177)(1317,174)(1337,171)(1357,168)
(1376,165)(1396,163)(1416,160)(1436,157)
\thinlines \path(485,832)(488,796)(493,750)(498,713)(503,684)(513,639)
(523,606)(533,580)(543,559)(563,526)(582,502)(602,482)(622,466)
(642,453)(662,441)(682,430)(702,421)(721,412)(741,404)(761,397)
(781,390)(801,384)(821,378)(840,372)(860,366)(880,361)(900,356)
(920,351)(940,346)(960,341)(979,337)(999,332)(1019,328)(1039,324)
(1059,320)(1079,315)(1099,311)(1118,308)(1138,304)(1158,300)(1178,296)
(1198,292)(1218,289)(1237,285)(1257,282)(1277,278)(1297,274)(1317,271)
(1337,268)(1357,264)(1376,261)
\thinlines \path(1376,261)(1396,257)(1416,254)(1436,251)
\thinlines \path(220,277)(220,277)(220,277)(220,278)(221,278)
(221,278)(221,278)(221,278)(222,279)(223,279)(223,279)(224,279)
(225,279)(226,279)(228,280)(229,280)(230,280)(231,280)(233,280)
(233,280)(234,280)(234,280)(235,280)(235,280)(236,280)(236,280)
(236,280)(236,280)(237,280)(237,280)(237,280)(237,280)(237,280)
(237,280)(237,280)(238,280)(238,280)(238,280)(238,280)(238,280)
(239,280)(239,280)(240,280)(241,280)(242,280)(245,280)(250,280)
(255,280)(260,279)(265,279)(270,279)
\thinlines \path(270,279)(275,279)(280,279)(285,278)(290,278)(294,278)
(299,278)(304,278)(309,278)(314,277)(319,277)(324,277)(329,277)(332,277)
(334,277)(337,277)(338,277)(339,277)(340,277)(341,277)(342,277)(342,277)
(342,277)(343,277)(343,277)(343,277)(343,277)(343,277)(343,277)(344,277)
(344,277)(344,277)(344,277)(344,277)(344,277)(345,277)(345,277)(345,277)
(345,277)(346,277)(347,277)(348,277)(349,277)(350,277)(352,277)(354,277)
(356,277)(359,277)(364,277)(369,278)(374,278)
\thinlines \path(374,278)(379,278)(384,278)(389,279)(394,280)(399,280)
(404,281)(409,282)(413,283)(418,285)(423,287)(428,289)(433,293)(438,297)
(441,300)(443,303)(446,307)(448,312)(451,318)(452,322)(453,327)(454,333)
(456,340)(457,349)(458,355)(458,362)(459,380)(460,392)(460,400)(461,409)
(461,420)(461,434)(462,451)(462,474)(462,488)(462,506)(462,528)(462,557)
(463,596)(463,655)(463,758)(463,832)
\thinlines \path(220,292)(220,292)(220,293)(220,294)(221,295)(221,296)
(221,296)(222,297)(223,298)(224,299)(225,300)(228,301)(230,302)(235,303)
(240,305)(245,306)(250,307)(255,307)(260,308)(265,309)(270,310)(275,310)
(280,311)(285,312)(290,313)(294,313)(299,314)(304,315)(309,316)(314,317)
(319,318)(324,319)(329,320)(334,321)(339,322)(344,324)(349,325)(354,326)
(359,328)(364,330)(369,332)(374,334)(379,336)(384,339)(389,342)(394,345)
(399,349)(404,353)(409,357)(413,363)(418,369)
\thinlines \path(418,369)(423,377)(428,386)(433,397)(438,412)(441,421)
(443,432)(446,445)(448,462)(451,483)(452,496)(453,511)(454,529)(456,552)
(457,582)(458,600)(458,621)(459,679)(460,719)(460,744)(461,774)(461,809)
(461,832)
\thinlines \path(220,336)(220,336)(220,340)(220,342)(221,345)(221,347)
(221,349)(223,354)(223,356)(224,358)(225,361)(230,369)(235,375)(240,380)
(245,385)(250,389)(255,393)(260,397)(265,401)(270,405)(275,408)(280,412)
(285,416)(290,420)(294,423)(299,427)(304,431)(309,435)(314,440)(319,444)
(324,449)(329,453)(334,458)(339,464)(344,469)(349,475)(354,481)(359,488)
(364,495)(369,502)(374,510)(379,519)(384,529)(389,539)(394,551)(399,563)
(404,578)(409,594)(413,612)(418,634)(423,659)
\thinlines \path(423,659)(428,689)(433,727)(438,775)(441,805)(443,832)
\end{picture}
\caption{Behavior of $\varphi_{\mbox{cr}}-\varphi$ as a function of 
$\varphi$ for
$\mu=2M_{pl}$ and $c<0$. The three typical cases of the gauge coupling 
strength are
shown: $-cg^2=10^{-3},10^{-2},10^{-1}$.}
\end{figure}
\pagebreak
\begin{figure}
\setlength{\unitlength}{0.240900pt}
\begin{picture}(1500,900)(0,0)
\tenrm
\thinlines \drawline[-50](220,273)(1436,273)
\thicklines \path(220,113)(240,113)
\thicklines \path(1436,113)(1416,113)
\put(198,113){\makebox(0,0)[r]{-10}}
\thicklines \path(220,193)(240,193)
\thicklines \path(1436,193)(1416,193)
\put(198,193){\makebox(0,0)[r]{-5}}
\thicklines \path(220,273)(240,273)
\thicklines \path(1436,273)(1416,273)
\put(198,273){\makebox(0,0)[r]{0}}
\thicklines \path(220,353)(240,353)
\thicklines \path(1436,353)(1416,353)
\put(198,353){\makebox(0,0)[r]{5}}
\thicklines \path(220,433)(240,433)
\thicklines \path(1436,433)(1416,433)
\put(198,433){\makebox(0,0)[r]{10}}
\thicklines \path(220,512)(240,512)
\thicklines \path(1436,512)(1416,512)
\put(198,512){\makebox(0,0)[r]{15}}
\thicklines \path(220,592)(240,592)
\thicklines \path(1436,592)(1416,592)
\put(198,592){\makebox(0,0)[r]{20}}
\thicklines \path(220,672)(240,672)
\thicklines \path(1436,672)(1416,672)
\put(198,672){\makebox(0,0)[r]{25}}
\thicklines \path(220,752)(240,752)
\thicklines \path(1436,752)(1416,752)
\put(198,752){\makebox(0,0)[r]{30}}
\thicklines \path(220,832)(240,832)
\thicklines \path(1436,832)(1416,832)
\put(198,832){\makebox(0,0)[r]{35}}
\thicklines \path(220,113)(220,133)
\thicklines \path(220,832)(220,812)
\put(220,68){\makebox(0,0){0}}
\thicklines \path(463,113)(463,133)
\thicklines \path(463,832)(463,812)
\put(463,68){\makebox(0,0){2}}
\thicklines \path(706,113)(706,133)
\thicklines \path(706,832)(706,812)
\put(706,68){\makebox(0,0){4}}
\thicklines \path(950,113)(950,133)
\thicklines \path(950,832)(950,812)
\put(950,68){\makebox(0,0){6}}
\thicklines \path(1193,113)(1193,133)
\thicklines \path(1193,832)(1193,812)
\put(1193,68){\makebox(0,0){8}}
\thicklines \path(1436,113)(1436,133)
\thicklines \path(1436,832)(1436,812)
\put(1436,68){\makebox(0,0){10}}
\thicklines \path(220,113)(1436,113)(1436,832)(220,832)(220,113)
\put(45,922){\makebox(0,0)[l]{\shortstack{$(\varphi_{cr}-\varphi)/M_{pl}$}}}
\put(828,-22){\makebox(0,0){$\varphi/M_{pl}$}}
\put(828,877){\makebox(0,0){$\mu/M_{pl} = 2$ , $c > 0$}}
\put(950,385){\makebox(0,0)[l]{$cg^2 = 10^{-3}$}}
\put(706,321){\makebox(0,0)[l]{$cg^2 = 10^{-2}$}}
\put(706,177){\makebox(0,0)[l]{$cg^2 = 10^{-1}$}}
\thinlines \path(464,832)(464,557)(465,474)(465,434)(466,410)(466,392)
(467,380)(468,361)(469,355)(470,349)(471,339)(472,332)(473,326)(476,317)
(478,310)(481,305)(483,300)(488,293)(493,288)(503,280)(513,274)(523,270)
(543,263)(563,257)(582,252)(602,248)(622,244)(642,240)(662,237)(682,233)
(702,230)(721,227)(741,224)(761,220)(781,217)(801,214)(821,211)(840,208)
(860,206)(880,203)(900,200)(920,197)(940,194)(960,191)(979,188)(999,186)
(1019,183)(1039,180)(1059,177)(1079,175)(1099,172)
\thinlines \path(1099,172)(1118,169)(1138,166)(1158,164)(1178,161)(1198,158)
(1218,155)(1237,153)(1257,150)(1277,147)(1297,145)(1317,142)(1337,139)
(1357,136)(1376,134)(1396,131)(1416,128)(1436,126)
\thinlines \path(465,832)(466,775)(466,721)(467,681)(468,624)(469,603)
(470,584)(471,555)(472,532)(473,514)(476,485)(478,464)(481,448)(483,435)
(488,414)(493,399)(498,387)(503,377)(513,362)(523,351)(543,334)(563,322)
(582,312)(602,304)(622,298)(642,291)(662,286)(682,281)(702,276)(721,272)
(741,267)(761,263)(781,259)(801,255)(821,252)(840,248)(860,244)(880,241)
(900,238)(920,234)(940,231)(960,228)(979,224)(999,221)(1019,218)
(1039,215)(1059,212)(1079,209)(1099,206)(1118,203)(1138,200)
\thinlines \path(1138,200)(1158,197)(1178,194)(1198,191)(1218,188)
(1237,185)(1257,182)(1277,179)(1297,176)(1317,173)(1337,170)(1357,168)
(1376,165)(1396,162)(1416,159)(1436,156)
\thinlines \path(485,832)(488,796)(493,750)(498,713)(503,684)(513,639)
(523,606)(533,580)(543,559)(563,526)(582,501)(602,482)(622,466)(642,452)
(662,441)(682,430)(702,421)(721,412)(741,404)(761,397)(781,390)(801,383)
(821,377)(840,372)(860,366)(880,361)(900,355)(920,351)(940,346)(960,341)
(979,336)(999,332)(1019,328)(1039,323)(1059,319)(1079,315)(1099,311)
(1118,307)(1138,303)(1158,300)(1178,296)(1198,292)(1218,289)(1237,285)
(1257,281)(1277,278)(1297,274)(1317,271)(1337,267)(1357,264)(1376,261)
\thinlines \path(1376,261)(1396,257)(1416,254)(1436,250)
\thinlines \path(220,282)(220,282)(220,282)(220,282)(221,282)(221,282)
(221,282)(221,282)(222,282)(223,283)(223,283)(224,283)(224,283)(225,283)
(226,283)(226,283)(227,283)(227,283)(228,283)(228,283)(228,283)(228,283)
(228,283)(229,283)(229,283)(229,283)(229,283)(229,283)(229,283)(230,283)
(230,283)(230,283)(230,283)(230,283)(231,283)(231,283)(232,283)(233,283)
(235,283)(237,283)(240,283)(245,282)(250,282)(255,282)(260,282)(265,281)
(270,281)(275,281)(280,281)(285,280)(290,280)
\thinlines \path(290,280)(294,280)(299,280)(304,280)(309,279)(314,279)
(319,279)(324,279)(329,279)(334,279)(337,279)(339,279)(342,278)(343,278)
(344,278)(345,278)(347,278)(347,278)(348,278)(348,278)(349,278)(349,278)
(349,278)(349,278)(350,278)(350,278)(350,278)(350,278)(350,278)(350,278)
(350,278)(351,278)(351,278)(351,278)(351,278)(352,278)(352,278)(353,278)
(353,278)(354,278)(355,278)(356,278)(359,279)(361,279)(364,279)(369,279)
(374,279)(379,279)(384,280)(389,280)(394,281)
\thinlines \path(394,281)(399,281)(404,282)(409,283)(413,284)(418,286)
(423,288)(428,290)(433,293)(438,297)(441,300)(443,303)(446,307)(448,312)
(451,319)(452,323)(453,327)(454,333)(456,340)(457,349)(458,355)(458,362)
(459,380)(460,393)(460,400)(461,410)(461,421)(461,434)(462,451)(462,474)
(462,489)(462,506)(462,528)(462,557)(463,596)(463,655)(463,758)(463,832)
\thinlines \path(220,294)(220,294)(220,295)(220,295)(221,296)(221,297)
(221,297)(222,298)(223,299)(224,300)(225,301)(228,302)(230,303)(235,304)
(240,306)(245,306)(250,307)(255,308)(260,309)(265,310)(270,310)(275,311)
(280,312)(285,313)(290,313)(294,314)(299,315)(304,316)(309,316)(314,317)
(319,318)(324,319)(329,320)(334,321)(339,323)(344,324)(349,325)(354,327)
(359,329)(364,330)(369,332)(374,334)(379,337)(384,339)(389,342)(394,345)
(399,349)(404,353)(409,358)(413,363)(418,369)
\thinlines \path(418,369)(423,377)(428,386)(433,397)(438,412)(441,421)
(443,432)(446,445)(448,462)(451,483)(452,496)(453,511)(454,530)(456,553)
(457,582)(458,600)(458,621)(459,679)(460,719)(460,744)(461,774)(461,809)
(461,832)
\thinlines \path(220,336)(220,336)(220,340)(220,342)(221,345)(221,348)
(221,349)(223,355)(223,356)(224,358)(225,361)(230,369)(235,376)(240,381)
(245,385)(250,390)(255,394)(260,397)(265,401)(270,405)(275,409)(280,412)
(285,416)(290,420)(294,424)(299,427)(304,431)(309,436)(314,440)(319,444)
(324,449)(329,454)(334,459)(339,464)(344,469)(349,475)(354,481)(359,488)
(364,495)(369,502)(374,511)(379,519)(384,529)(389,539)(394,551)(399,564)
(404,578)(409,594)(413,612)(418,634)(423,659)
\thinlines \path(423,659)(428,689)(433,727)(438,775)(441,805)(443,832)
\end{picture}
\caption{Behavior of $\varphi_{\mbox{cr}}-\varphi$ as a function of 
$\varphi$ for
$\mu=2M_{pl}$ and $c>0$. The three typical cases of the gauge coupling 
strength are
shown: $cg^2=10^{-3},10^{-2},10^{-1}$.}
\end{figure}
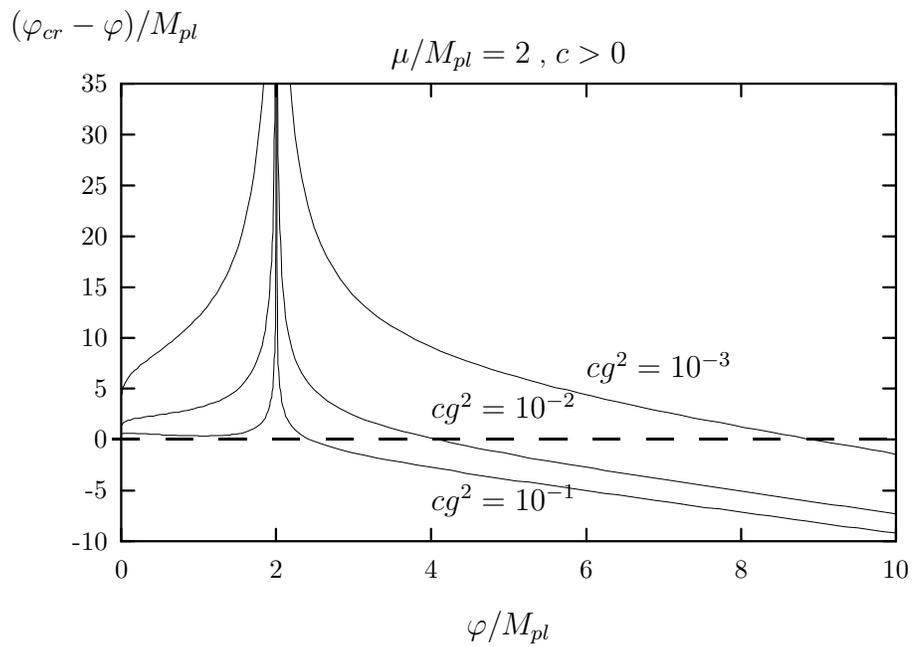
\end{document}